\documentclass[aps,prx,twocolumn,superscriptaddress,nofootinbib,floatfix]{revtex4-2}
\usepackage{amsmath,amssymb,wasysym,graphicx}
\usepackage{times}
\usepackage[varg]{txfonts}
\usepackage{textcomp}
\usepackage{subfigure}
\usepackage{tabu}
\usepackage{color}
\usepackage[colorlinks=true,citecolor=blue,urlcolor=blue,linkcolor=blue,hyperindex]{hyperref}
\usepackage{braket}
\usepackage{overpic}
\usepackage{appendix}

\usepackage[normalem]{ulem}
\usepackage{verbatim}
\allowdisplaybreaks

\begin{document}
\title{Evolution of dynamical signature in the X-cube fracton topological order}
\author{Chengkang Zhou}
\thanks{These authors contributed equally to this work.}
\affiliation{Department of Physics and HKU-UCAS Joint Institute of Theoretical and Computational Physics,The University of Hong Kong, Pokfulam Road, Hong Kong SAR, China}
\author{Meng-Yuan Li}
\thanks{These authors contributed equally to this work.}
\affiliation{School of Physics, State Key Laboratory of Optoelectronic Materials and Technologies, and Guangdong Provincial Key Laboratory of Magnetoelectric Physics and Devices, Sun Yat-sen University, Guangzhou, 510275, China}
\author{Zheng Yan}\email{zhengyan@hku.hk}
\affiliation{Department of Physics and HKU-UCAS Joint Institute of Theoretical and Computational Physics,The University of Hong Kong, Pokfulam Road, Hong Kong SAR, China}
\affiliation{State Key Laboratory of Surface Physics and Department of Physics, Fudan University, Shanghai 200438, China}
\author{Peng Ye}\email{yepeng5@mail.sysu.edu.cn}
\affiliation{School of Physics, State Key Laboratory of Optoelectronic Materials and Technologies, and Guangdong Provincial Key Laboratory of Magnetoelectric Physics and Devices, Sun Yat-sen University, Guangzhou, 510275, China}
\author{Zi Yang Meng}
\affiliation{Department of Physics and HKU-UCAS Joint Institute of Theoretical and Computational Physics,The University of Hong Kong, Pokfulam Road, Hong Kong SAR, China}

\begin{abstract}
As an unconventional realization of topological orders with an exotic interplay of topology and geometry, fracton (topological) orders feature subextensive topological ground state degeneracy and subdimensional excitations that are movable only within a certain subspace. It has been known in the exactly solvable three-dimensional X-cube model that universally represents the type-I fracton orders, that mobility constraints on subdimensional excitations originate from the absence of spatially deformable string-like operators. To unveil the interplay of topology and geometry, in this paper, we study the dynamical signature in the X-cube model in the presence of external Zeeman  fields via large-scale quantum Monte Carlo simulation and stochastic analytic continuation. We compute both real-space correlation functions and dynamic structure factors of subdimensional excitations (i.e., fractons, lineons, and planons) in the fracton phase and their evolution into the trivial paramagnetic phase by increasing external fields. We find in the fracton phase, that the correlation functions and the spectral functions show clear anisotropy exactly caused by  the underlying mobility constraints. On the other hand, the external fields successfully induce quantum fluctuations and offer mobility to excitations along  the subspace allowed by mobility constraints. These numerical results provide the evolution of a dynamical signature of subdimensional particles in fracton orders, indicating that the mobility constraints on local dynamical properties of subdimensional excitations are deeply related to the existence of fracton topological order.  The results will also be helpful in  potential experimental identifications in spectroscopy measurements such as neutron scattering and nuclear magnetic resonance.
\end{abstract}
\date{\today}
	
\maketitle

\section{Introduction}
\label{sec:Introduction}
The exploration of novel phases of matter beyond the Landau symmetry breaking paradigm is one of the major themes in modern condensed matter physics. Within numerous important findings, topological orders that feature long-range entanglement and their experimental detection in correlated materials and numerical simulation in quantum lattice models, have attracted a lot of attention~\cite{Wen2019,Levin2007,kitaev2003fault,Levin2004,Wen1998,wen1995topological,Kitaev2009,Dynamical2018Sun,topological2021yan,wang2021fractionalized,wang2021fractionalized,Quantum2018Wang,YiZhou2017,Broholm2020,ZY2021ES,JRZhao2022}. Topological excitations are one of the defining features of topological orders, such as anyons in the celebrated toric code model~\cite{kitaev2003fault} and its frustrated magnet realizations~\cite{YCWang2017QSL,Dynamical2018Sun,Quantum2018Wang,YCWang2021vestigial,YCWang2021,isakov2011topological,zhou2022quantum,isakov2012FractionalizedQCP,ZY2022vd,ZY2022Loop}. While a single topologically trivial excitation can be created by a local operator, point like topological excitations, must be created in pairs, which are located at the two endpoints of string-like operators. As endpoints are exactly the boundaries of open strings, the topological excitations are inherently related to string operators. %That is to say, such topological excitations are insensitive to the geometry of string operators.
Since the string operators can be arbitrarily deformed, the associated topological excitations can move freely in the whole space.

Recently, fracton orders as an unconventional realization of topological orders were proposed and have stimulated intensive activities in many areas ranging from condensed matter physics and quantum information science,
 %\cite{qm2016Rev,CSS1996Shor}
to high energy physics~\cite{Chamon05fracton, Vijay2015,Fracton2016Vijay,Haah2011, Yoshida13, Fracton2016Vijay,Fractons2019Nandkishore,qm2016Rev,2020Fracton,Hermele17, Vijay2015,Vijay17coupledLayersXcube, Chen20coupledCS, Aasen20tqftnetwork, Slagle21foliatedfield, XuWu08RVP,Pretko17tensorgauge, Chen18tensorhiggs, Barkeshli18tensorhiggs, Seiberg21ZNfracton,Pretko18fractongaugeprinciple, Gromov19multipoleAlgebra, Seiberg19vectorSym,Shirley2018,LiYe20fracton, LiYe21fracton,2020PhRvR2b3267Y,Chen2021FS,Ye2021PRR,Yuan:2022mns,Zhu:2022lbx,2022arXiv220306984Y,helmes2017entanglement,Correlation2018Devakul,Quantum2020Muhlhauser,Nandkishore21sym,Foliated2019Kevin,Prem2017,jain_xu_ryd2022}. Exactly solvable models that support fracton orders (phases) are usually stabilizer codes defined in three and higher dimensions. In sharp contrast to the conventional topological orders, locally-indistinguishable ground state degeneracy (GSD) in fracton system grows with the system size, such interesting phenomena can be traced back to the foliation structure built in the lattice models of three~\cite{Shirley2018} and higher dimensions~\cite{LiYe20fracton,LiYe21fracton}. Furthermore, while topological excitations in higher-dimensional conventional topological orders are characterized by topological properties, e.g., topological field theoretical description and topological braiding data among particles and loop-like excitations~\cite{lantian3dto1,lantian3dto2,yp18prl,ypdw,wang_levin1,bti2,jian_qi_14,string5,PhysRevX.6.021015,string6,ye16a,YeGu2015,corbodism3,YW13a,ye16_set,2018arXiv180101638N,2016arXiv161008645Y,string4,PhysRevLett.114.031601,3loop_ryu,string10,2016arXiv161209298P,Ye:2017aa,Tiwari:2016aa,2012FrPhy...7..150W,zhang2021compatible,Zhang:2021ycl},  topological excitations in fracton orders are unexpectedly exotic   due to mobility constraints, which can be seen from the fact that spatially deformable string operators do not exist~\cite{Haah2011} in exactly solvable fracton models, e.g., the X-cube model~\cite{Fracton2016Vijay} and the Haah's code~\cite{Haah2011} in three-dimensional (3D) lattice. In the literature, topological excitations in fracton orders are dubbed \textit{fractons} and \textit{subdimensional particles} according to the degree of their mobility~\cite{Vijay2015}, with fracton denoting excitations that are completely immobile ($0$-dimensional), and subdimensional particle denoting excitations that are still mobile within a subspace including $1$-dimensional \textit{lineons} moving along straight lines (e.g., straight lines formed by nearest links along the $x$ direction) and $2$-dimensional \textit{planons} moving within flat planes (e.g., all flat planes formed by nearest plaquettes that are parallel to $xy$ planes). In this paper, for the sake of convenience, we also include fractons as subdimensional particles. Interestingly, the spirit of this nomenclature has been generalized to higher dimensions, where spatially extended excitations are subject to complicated constraints on both mobility and deformability~\cite{LiYe20fracton,LiYe21fracton}. In addition, as a common feature of stabilizer codes, all these exactly solvable fracton models have exactly zero correlation length and trivial dynamics in the sense that all subdimensional excitations are static, dispersionless and localized energy lumps above the uniform ground-state energy background.

 In the fracton phase,   mobility constraints, which are geometric (local dynamical) properties, have topological origin and can be utilized to universally characterize the phase in addition to size-dependent formula of the subextensive GSD. In exactly solvable models, such as the X-cube model and the Haah's code model, the origin of restricted mobility of subdimensional particles can be clearly demonstrated. However, the situation has been unclear in general lattice models that lose exact solvability. To understand fracton phases on a more general ground, it is highly desirable to find a model family, such that a well-known fracton model, e.g., the X-cube model can be continuously deformed away from the exactly solvable point and eventually evolves into a trivial phase via quantum phase transition.
 Interestingly, it has been known that in the toric code model that is a simple stabilizer code model, anyons gain dispersive dynamics when external Zeeman couplings is tuned on, which renders fascinating anyon condensation of the celebrated Bais-Slingerland mechanism~\cite{bais2009} at the quantum critical points between the $\mathbb{Z}_2$ topological order and trivial phases in 2D~\cite{YCWang2017QSL,Dynamical2018Sun,Quantum2018Wang,YCWang2021vestigial,YCWang2021,isakov2011topological,isakov2012FractionalizedQCP,Kamiya2015Magnetic,Robustness2013Saeed,ac0,ac1,ac2,ac3,ac4,ac5,ye10,ac6}. Recently such a Bais-Slingerland-like mechanism in   more general topological phases has also been generalized to 3D~\cite{ye16a,YW13a,bti2,YeGu2015}.  For the purpose of the present work, the model family, i.e., ``stabilizer code + Zeeman fields'' is intriguing~\cite{Universality2019Zack,CSS1996Shor,qm2016Rev,Zhu19ToricCode,reiss2019quantum} both theoretically and numerically, which offers an opportunity to think, in a more practical way, the characterization of topological phases in a large parameter space. Along this line of thinking, in this paper, we aim to study fracton orders when the 3D lattice models are driven away from exactly solvable points by turning on Zeeman couplings and the quantum critical points may be signaled by condensations of subdimensional excitations. We will especially focus on the evolution of dynamical signature of subdimensional particles, which is still lacking in the literature. We expect that dynamical signature will encode fingerprint of fracton orders from spectroscopic probes and exhibit the nontrivial interplay of topology and geometry in such unconventional realization of topological orders.

The numerical simulation of fracton system is notoriously difficult and arduous, and we have improved the calculation methods and overcome the difficulties in various simulations. By means of quantum Monte Carlo (QMC) and stochastic analytic continuation (SAC)~\cite{sandvik2010computational,Stochastic2003Sandvik,yan2020improved,CKZhou2021,AWS2022process,Dynamical2018nvsen,Dynamical2018Sun,Sandvik1998Stochastic,beach2004identifying,Using2008Syljuasen}, we consider the X-cube model  in the presence of Zeeman fields and numerically study dynamical signature of subdimensional particles, i.e., fractons, lineons, and planons. The reason of choosing the X-cube model is that this model, from the generalized local unitary transformation point of view, captures   universal properties of the so-called type-I fracton order~\cite{Shirley2018}. %which we refer as "the X-cube fracton order'' throughout the paper.
 In the literature, several investigations on both analytical and numerical sides have been performed~\cite{helmes2017entanglement,Correlation2018Devakul,Quantum2020Muhlhauser,Zhu:2022lbx}. For instance, the classical variant of the X-cube model is studied and thermal properties such as a moving peak in its specific heat due to the finite-size have been found~\cite{helmes2017entanglement}. Besides, with the help of QMC simulation and the perturbative continuous unitary transformation (pCUT) method, the ground state phase transitions of the X-cube model under the single field perturbation have been confirmed to be first-order~\cite{Correlation2018Devakul,Quantum2020Muhlhauser}, which is tightly connected to immobility nature of fracton excitations. The nature of the transitions is also systematically studied by generalizing the $\mathbb{Z}_2$ X-cube model to its $\mathbb{Z}_N$ cousins in terms of the tensor network state representations~\cite{Zhu:2022lbx}. Moreover, dispersion relations of single quasiparticle excitations are perturbatively computed~\cite{Quantum2020Muhlhauser}.

 Here, with the QMC+SAC scheme, we systematically compute real-space density-density correlation functions and frequency-momentum-dependence of spectral functions (dynamical structure factors) of subdimensional particles. From the QMC numerical results, we find that   mobility constraints on subdimensional particles deeply influence {real-space correlation functions} and  {spectral functions} in many aspects. Starting from the zero field limit where all subdimensional particles are dispersonless,
 our numerical results indicate that,   in the presence of quantum fluctuations generated by the Zeeman fields,  the subdimensional particles acquire dispersive dynamics \textit{strictly} under   mobility constraints.  After the quantum phase transition where the fracton phase is turned into the trivial paramagnetic phase, the anisotropic features under   mobility constraints   disappear.
In addition to QMC simulation, we also perform a mean-field+RPA (random phase approximation) analysis to fit the QMC results of spectral functions, which shows a qualitative consistency with the QMC results.

Our QMC results provide a set of dynamical signature of subdimensional particles in the type-I fracton orders, which clearly demonstrates an exotic interplay of topology and symmetry in fracton physics and shows sharp distinction from the conventional topological orders.  As spectroscopy measurements, e.g.,   inelastic neutron scattering and nuclear magnetic resonance, play important roles in identifying novel phases of matter in  strongly correlated materials, we expect the QMC results in the present work will be helpful in experimentally identifying fracton orders in real materials. In addition, the QMC results also present  various features in correlation functions and spectral functions, which are yet to be fully understood, so we  expect  our numerical study will stimulate further studies in the field of fracton physics from both theoretical and numerical aspects.

This paper is organized as follows. In Sec.~\ref{sec:Model}, we introduce the theoretical background of the X-cube model and the density operator of three different subdimensional excitations: fracton, lineon and planon. Theoretical    expectation  on  correlation functions and dynamical structure factors is given in Sec.~\ref{sec:theo_pre}. In Sec.~\ref{sec:numerical calculation},   QMC+SAC results are presented in details, where the comparison between QMC and meanfield+RPA analysis is also provided.
%The detail discussion are given in Sec.~\ref{sec:discussion}.
We summarize the paper in Sec.~\ref{sec:discussion}.   Appendix~\ref{sec:numerical method} contains several technical details of our numerical method.

\section{The X-cube model in the presence of Zeeman fields}
\label{sec:XCModel}
\subsection{Model Hamiltonian and subdimensional excitations}\label{sec:Model}

The Hamiltonian of the X-cube model with external Zeeman-like magnetic fields is expressed as~\cite{Fracton2016Vijay,Correlation2018Devakul,Quantum2020Muhlhauser,Foliated2019Kevin,helmes2017entanglement}
\begin{equation}
	\begin{aligned}
		H&=-K\sum_{i}A_{c,i}-\Gamma\sum_{i,v}B_{v,i}-h_{x}\sum_{i}\sigma^{x}_{i}-h_{z}\sum_{i} \sigma^{z}_{i}.
		\label{eq:eq1}
	\end{aligned}
\end{equation}

\begin{figure}[htp!]
	\centering
	\includegraphics[width=\columnwidth]{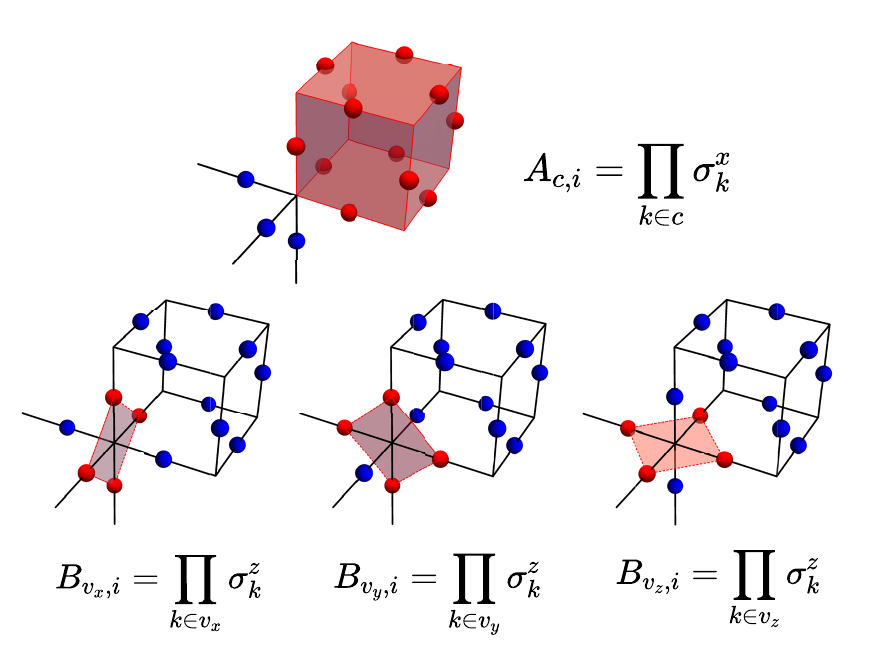}
	\caption{An illustration of the stabilizer operators in the X-cube model. $A_{c,i}$ denotes the product of all $\sigma_{x}$ in the each cube (the red shadowed cube), while $B_{v,i}$ denotes the product of $\sigma_{z}$ in the each cross (the red shadowed planes).}
	\label{fig:fig1}
\end{figure}
As shown in Fig.~\ref{fig:fig1}, spin-$\frac{1}{2}$s are defined on links, and $A_{c,i}$ ($B_{v,i}$) denotes the product of all the $\sigma_{x}$ ($\sigma_{z}$) in the each cube (cross), respectively~\cite{Fracton2016Vijay,Fractons2019Nandkishore,Foliated2019Kevin,Cage-Net2019Prem,isotropic2017vijay,helmes2017entanglement}. Here the summation index $i$ in general refers to a spatial location, while its specific meaning depends on the summed operator, for example, in $A_{c,i}$($B_{v,i}$, $\sigma^{z}_{i}$) $i$ refers to the coordinates of the center of a cube(vertex, link). As in this paper we mainly consider the distance between operators of the same kind, this ambiguity should not lead to any misunderstanding. There are three different kinds of subdimensional excitations in the X-cube model, which are lineon, fracton, and planon. A lineon corresponds to the flipped eigenvalues of two different $B_{v,i}$ terms at the same site, which combinatorially leads to three different species of itself. And a fracton corresponds to the flipped eigenvalue of the cube operator $A_{c,i}$, which are completely immobile. A pair of fractons along the $x$ ($y$, $z$) direction correspond to a planon which can move in the 2D plane perpendicular to $x$ ($y$, $z$) direction.

\begin{figure}[htp!]
	\centering
	\includegraphics[width=\columnwidth]{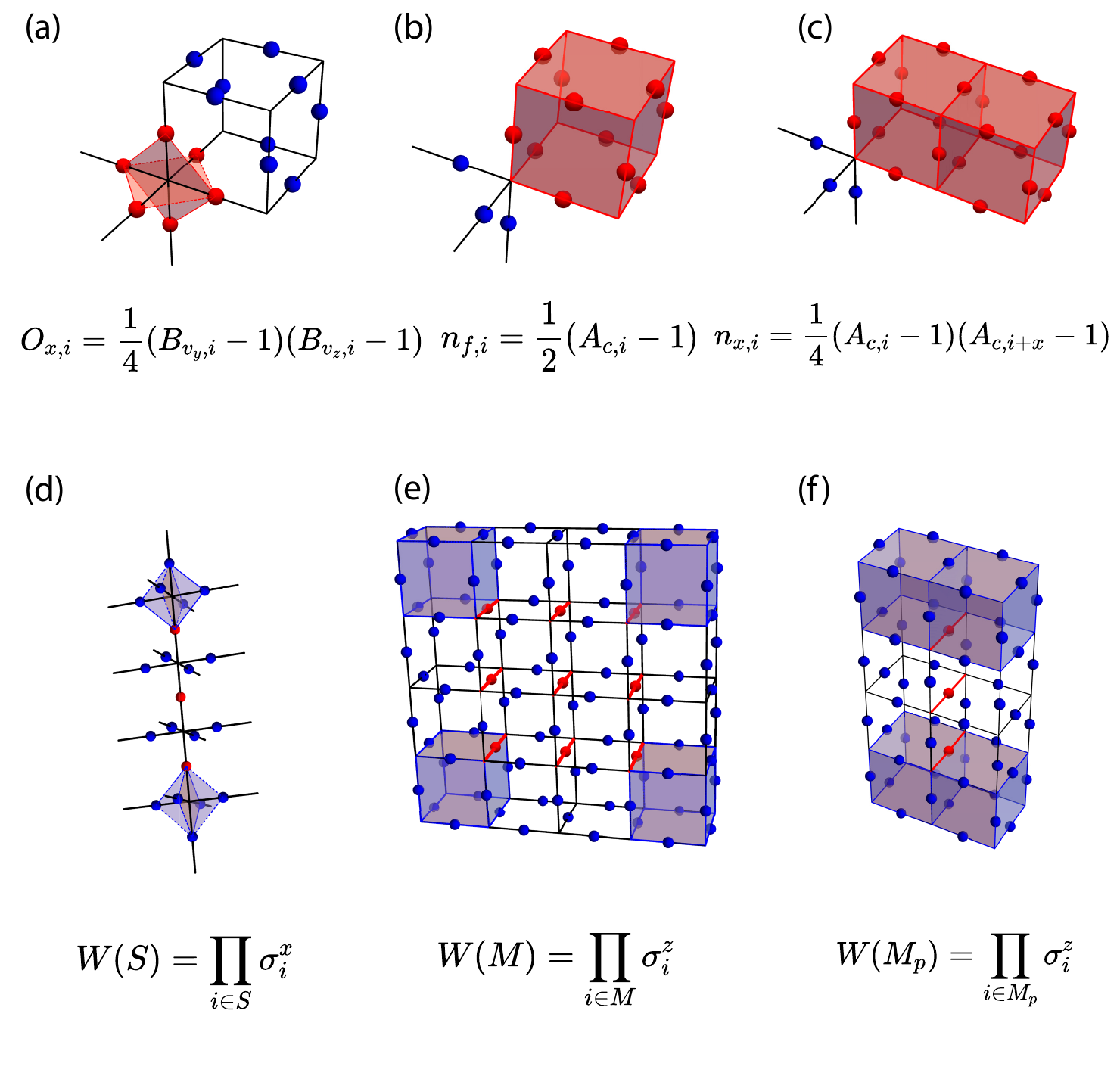}
	\caption{An illustration of the density operators of various subdimensional excitations in the X-cube model. (a), (b), and (c) are respectively for lineons, fractons, and planons. And (d), (e), and (f) are the creation operators of lineons, fractons, and planons, where the red spheres are the spin contained in the creation operators.}
	\label{fig:fig2}
\end{figure}

We can obtain   mobility constraints by considering the creation operators of excitations. First, fractons are generated by membrane operators of the form $W(M) = \prod_{i\in M} \sigma_i^z$, where $M$ is a square membrane composed of parallel links, $i\in M$ means that link $i$ within $M$ (see Fig.\ref{fig:fig2}(e)). Such a membrane operator generates $4$ fractons respectively located at the four corners. Therefore, moving such a fracton with another membrane operator (i.e. apply a membrane operator to annihilate the original fracton, and create another one at another site) will generate $2$ extra fractons, that costs additional energy. Meanwhile, a planon, as a bound state of two fractons, is mobile along a two-dimensional manifold (see Fig.~\ref{fig:fig2}(f)). As for lineons, that are created by string operators of the form $W(S) = \prod_{i\in S} \sigma_i^x$. Here, $S$ is a straight string composed of links (see Fig.\ref{fig:fig2}(d)). Similar to fractons, moving a lineon at an endpoint of $S$ with another string operator perpendicular to $S$ will create an extra lineon at the intersection point of the two strings. Equivalently, we can say that for a $W(S^{'}) = \prod_{i\in S^{'}} \sigma_i^x$ operator, where $S^{'}$ is an arbitrary string, lineons are not only generated at its endpoints, but also its turning points. In brief, any movement perpendicular to the direction of $S$ is energetically unfavorable.

To investigate all these excitations in the QMC+SAC scheme~\cite{Dynamical2018Sun,Dynamical2018nvsen,Dynamics2019wei,Nearly2017Shao,sandvik2010computational,topological2021yan}, we compute the real-space correlation and their spectra (obtained from the SAC of dynamical correlation functions) in the X-cube model in both $\sigma_{x}$ and $\sigma_{z}$ basis with linear system size both $L=6$ and $L=10$ (the total size is $N=L^3$) and $\beta=2L$. For the dynamic correlation, the imaginary-time correlation functions in the lineon channel (Fig.~\ref{fig:fig2}(a))
\begin{equation} G_{O_{x}}(\textbf{q},\tau)=\frac{1}{L^3}\sum_{i,j}e^{-i\mathbf{q}\cdot(\mathbf{r}_{i}-\mathbf{r}_{j})}\langle O_{x,i}(\tau)O_{x,j}(0) \rangle
	\label{eq:eqOx}
\end{equation}
with $O_{x,i}=\frac{1}{4}(B_{v_y,i}-1)(B_{v_z,i}-1)$, where lineons are created by $W(S) = \prod_{i\in S} \sigma_i^x$ (Fig.~\ref{fig:fig2}(d));
the fracton channel (Fig.~\ref{fig:fig2}(b))
\begin{equation} G_{n_{f}}(\textbf{q},\tau)=\frac{1}{L^3}\sum_{i,j}e^{-i\mathbf{q}\cdot(\mathbf{r}_{i}-\mathbf{r}_{j})}\langle n_{f,i}(\tau)n_{f,j}(0) \rangle
\label{eq:eqnf}
\end{equation}
with $n_{f,i}=\frac{1}{2}(A_{c,i}-1)$, where fractons are created by $W(M) = \prod_{i\in M} \sigma_i^z$ (Fig.~\ref{fig:fig2}(e)),
and the planon channel (Fig.~\ref{fig:fig2}(c))
\begin{equation} G_{n_{x}}(\textbf{q},\tau)=\frac{1}{L^3}\sum_{i,j}e^{-i\mathbf{q}\cdot(\mathbf{r}_{i}-\mathbf{r}_{j})}\langle n_{x,i}(\tau)n_{x,j}(0) \rangle
\label{eq:eqnx}
\end{equation}
with $n_{x,i}=\frac{1}{4}(A_{c,i}-1)(A_{c,i+x}-1)$, where planons are created by $W(M_{p}) = \prod_{i\in M_{p}} \sigma_i^z$ (Fig.~\ref{fig:fig2}(f)),, are measured in the QMC simulation. Here, operators $O_x$, $n_{f}$ and $n_{x}$ denote the number operator of lineon, fracton and planon~\cite{qm2016Rev}, respectively. Since the ground state by definition should be vacua of all excitations, one can verify that the expectation values of $n_f$,  $O_x$ and $n_x$ indeed vanish in the X-cube model's ground state. However, under the perturbation of Zeeman couplings $h_{z}$ and $h_{x}$, their expectation values would become non-zero caused by noncommutation between $A_{c,i}$ (or $B_{v,i}$) and Zeeman fields, which corresponds to the emergence of subdimensional excitations (see also Appendix.~\ref{sec:appA3}).

\subsection{Correlation and spectral functions influenced by  mobility constraints}
\label{sec:theo_pre}

Theoretically, we expect correlation functions to be an important way to characterize the fracton order of the X-cube model with perturbative external field. To see this point, we can compare the X-cube model with 3D toric code model (TCM)~\cite{kitaev2003fault,string2,Double_Claudio2008}. As a model with pure topological order, in 3D TCM, a pair of point-like charge excitations is generated at the $2$ endpoints of a string operator, and a loop-like flux excitation is generated on the boundary of an open membrane operator. That is to say, the relation between creation operators and excitations in 3D TCM is completely topological: excitations are always generated at the boundaries of creation operators. While as a fracton order model, in the X-cube model, as discussed in Sec.~\ref{sec:Model}, there is no such completely topological relation between excitations and creation operators. For example, in the X-cube model, fractons are generated at the corners of a membrane, and lineon are not only generated at the endpoints, but also at the turning points of a string. As a result, fracton orders show a complicated interplay between topology and geometry. Therefore, even though mobility restriction is resulted from the fracton order of the whole system, the correlation function should also contain information of mobility restriction, and such restriction would have intricate ramifications into the three types of subdimensional excitations.

First, we consider lineon excitations. The lineon-lineon correlation at long distances is expected to be anisotropic due to the one-dimensional mobility of a lineon. In real space, for lineons mobile along the direction denoted as $x$, the lineon-lineon correlation $\langle O_{x,i}(\tau)O_{x,j}(0) \rangle$ should decay more slowly along the mobile $x$-direction. In momentum space, we expect such anisotropy of correlation function $G_{O_{x}}(\textbf{q},\tau)$ to be present around the $\Gamma$ point, as a result of the mobility restriction.
Then, we consider planon excitations. As demonstrated in Sec.~\ref{sec:Model}, a planon is composed of a dipole of adjacent fracton excitations. There is no clear evidence that guarantees  an attractive interaction between adjacent fractons. Thus, whether or not the composite is a well-defined point-like excitation at long wavenlengths is hard to be confirmed. Due to the existence of intrinsic dipole-like structure of a planon, planon-planon correlation function should be anisotropic not only in the fracton ordered phase, but also in the paramagnetic phase with strong external field $h_z$. Hence, anisotropy of planon-planon correlation is not necessarily related to fracton orders.
%Besides, as in the X-cube model with external magnetic field there is no attractive interaction between fractons, the physical interpretation of planon-planon correlation would not results into bound state, as we will show in the numerics.
At last, we consider fracton excitations. As a fracton is totally immobile, the fracton-fracton correlation function at long wavelengths is expected to be isotropic. Therefore, fracton-fracton correlation is relatively simple, and we do not expect to read much information of fracton orders from such correlation functions. In addition to the correlation function, we can also expect the spectral functions of subdimensional particles to show dispersive behavior along their movable directions.

As the X-cube model itself is isotropic, the anisotropy of lineon-lineon correlation at long wavelengths suggests a hidden rotational symmetry breaking may be related to the type-I fracton order. As a concrete example, the X-cube model is obviously invariant under a $90^{\circ}$ spatial rotation with respect to $x$-axis, while the correlation functions of lineons mobile along $y$ (and $z$) directions do not have this symmetry. In contrast, for pure topological orders, such as the isotropic 3D TCM, we expect the charge-charge correlation to be also isotropic. Therefore, such anisotropy of lineon-lineon correlation (or an isotropy breaking) may be recognized as a signature of Type-I fracton orders. Nevertheless, in type-II fracton orders, such as the Haah's code~\cite{Fractons2019Nandkishore,Haah2013,Haah2011}, we do not expect to find such spatial symmetry breaking due to the lack of subdimensional mobile excitations.

\section{numerical resutls and analysis}
\label{sec:numerical calculation}
In this section, we present our QMC+SAC results of lineon, fracton, and planon excitations in the X-cube model with external Zeeman fields.

\subsection{Numerical setting and critical fields}
Numerical simulation of fracton systems is notoriously difficult and arduous. We first begin with a few technique details which are related with the fracton physics. We implement our QMC simulation in the framework of stochastic series expansion (SSE)~\cite{sandvik2010computational}, where the sampling configuration is constructed by Taylor expansion of the partition function in a chosen basis. In our case, it is $\{\sigma_{z}\}$ for lineon and $\{\sigma_{x}\}$ for fracton in the simulation. Unfortunately, the four ($B_{v,i}$) and twelve-spin ($A_{c,i}$) interactions in the X-cube model cause a glassy and fragmental configuration space, which leads to a low efficiency in the Monte Carlo sampling. For example, on one hand, the local update is not efficient for QMC simulation of the X-cube model, due to the extended interactions in the Hamiltonian. On the other hand, naively using the cluster update algorithm~\cite{Stochastic2003Sandvik,yan2019}, the cluster would rapidly extend due to the $B_{v,i}$ and $A_{c,i}$ interactions and an extremely large area of spin is always suggested to be flipped, but flipping these extremely large areas has the same effect as flipping few spins. To solve this problem, we firstly modify the cluster update algorithm to slow down the cluster growth, and apply it along with the local update in our simulation. Secondly, for system size $L=10$, we warm up our Markov chains from an initial configuration from an equilibrium configuration in $L=2$ QMC simulation (spatially repeated from $L=2$ to $L=10$), to help the larger size simulation to thermalize quickly. The details of our QMC update scheme are given in Appendix~\ref{sec:appA1},~\ref{sec:appA2} and ~\ref{sec:appA3}. Our numerical simulations are carried out under the periodic boundary condition with linear system size $L=6$ and $L=10$, and total size $N=L^3$ and inverse temperature $\beta=2L$.

\begin{figure}[htp!]
\centering
\includegraphics[width=\columnwidth]{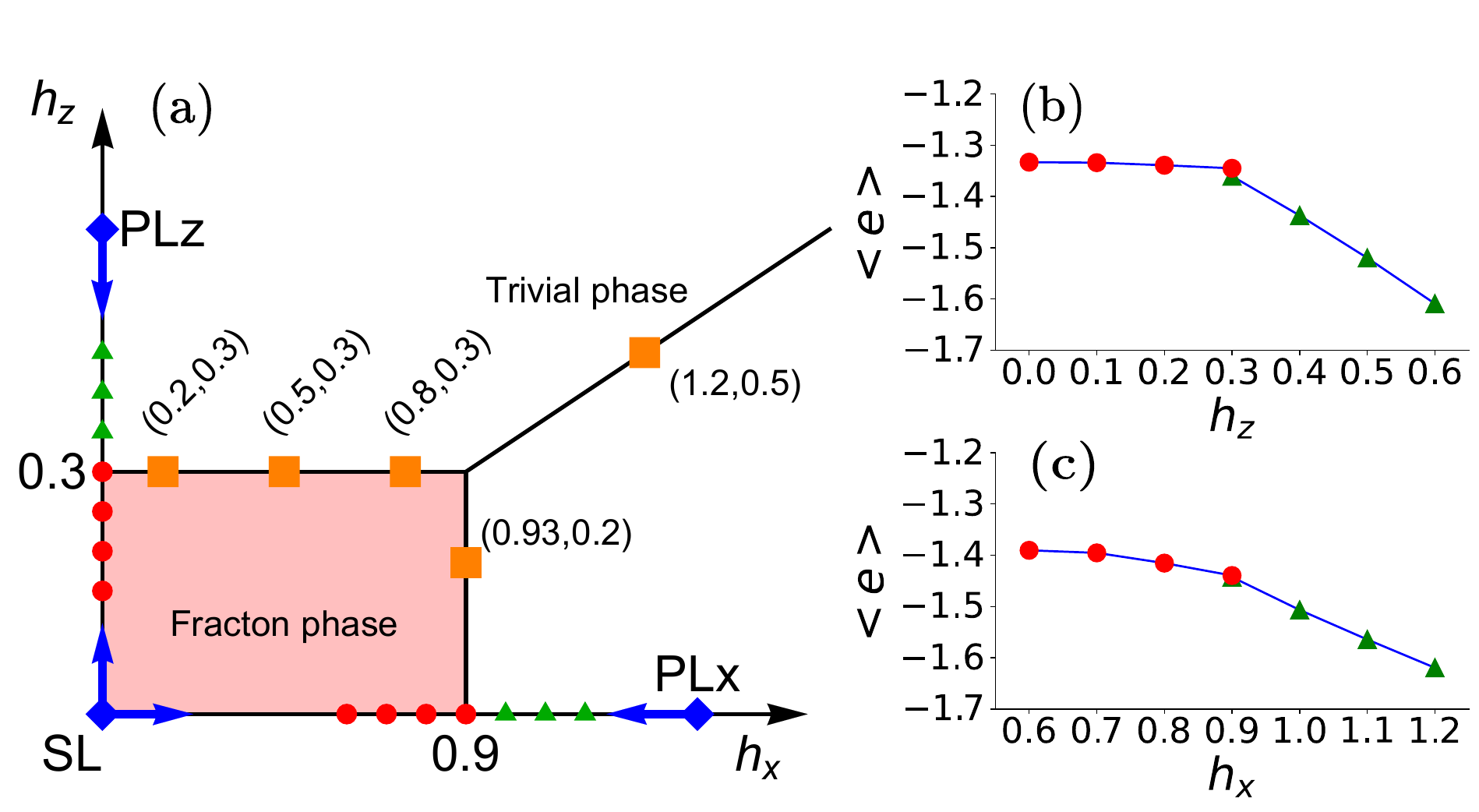}
\caption{Phase diagram of the X-cube model in the presence of external Zeeman fields. The blue arrows describe the scanning direction of QA in the QMC simulation. And the orange squares in (a) are the quantum critical points observed in in Fig.1 of the supplementary material in Ref.\cite{Correlation2018Devakul}. In the fracton phase, our simulation scans from the exactly solvable point [$(h_{x}=0,h_{z}=0)$, the blue diamond point]. And in the paramagnetic phase, we scan the X-cube model from the paramagnetic limit point in $x$-direction [PLx, $(h_{x}=2, h_{z}=0)$, the blue diamond point] and in $z$-direction [PLz, $(h_{x}=0,h_{z}=1)$, the blue diamond point]. Figures (b) and (c) are the energy per spin under the external fields with $L=10$. The red circles are measured by scanning from the exactly solvable point while the green triangles are from the PLz(x) points.}
\label{fig:fig3}
\end{figure}

Note that the ground state of the X-cube model is a vacuum state of all three subdimensional excitations, which means $O_{x}$, $n_{f}$ and $n_{x}$ all commute with the X-cube Hamiltonian Eq.~(\ref{eq:eq1}) with $h_{x}=0$ and $h_{z}=0$. Therefore, to observed these subdimensional excitations, the external perturbation fields are required to introduce a fluctuation to the corresponding operator (like $O_{x}$ for lineons). Here, we apply the $h_{x}$ transverse field to see lineon excitation and $h_{z}$ for fractons and planons. As described in Fig.~\ref{fig:fig3}(a), such perturbations also lead to a first-order phase transition from the fracton phase to the paramagnetic phase, which can be viewed as a consequence of the immobility of fracton excitation~\cite{Correlation2018Devakul,Quantum2020Muhlhauser,Zhu:2022lbx}. To compute correlations of these excitations with better data quality, beside the improvement in the QMC update scheme discussed above, we further utilize the quantum annealing (QA) algorithm \cite{kadowaki1998quantum,Car2002,yan2022targeting}, in which quantum parameter, $h_{x(z)}$, is slowly changed with the annealing step $\Delta h_{x(z)}=0.01$ for a faster convergence to the optimal state. In the fracton phase, starting from the exactly solvable point $(h_x=0, h_z=0)$, our simulations scan the parameter towards $(h_{x}=0.9, h_z=0)$ for the lineon and $(h_{z}=0.3, h_x=0)$ for the fracton cases, where we apply $2\times10^5$ Monte Carlo step for each annealing step. And for measurements in the paramagnetic phase, our simulations scan from the paramagnetic limit in $x$($z$)-direction [PLx(z)] point towards $h_{x}=0.9$ for the $h_{x}$ case and $h_{z}=0.3$ for the $h_{z}$. The detailed QA implementation is given in Appendix.~\ref{sec:appA3}. In Fig.~\ref{fig:fig3}(b,c), the energy per spin $\langle e\rangle$ are measured in both the fracton phase and the paramagnetic phase with $h_{x}$ and $h_{z}$ field on a $L=10$ system, respectively. These first-order phase transitions occur at $h_{x}\approx0.9$ in the $h_{x}$ perturbing case and $h_{z}\approx0.3$ for the $h_{z}$ case.

%\section{numerical results}
%\label{sec:numerical results}
\subsection{Correlation functions}
\begin{figure}[htp!]
\centering
	\includegraphics[width=\columnwidth]{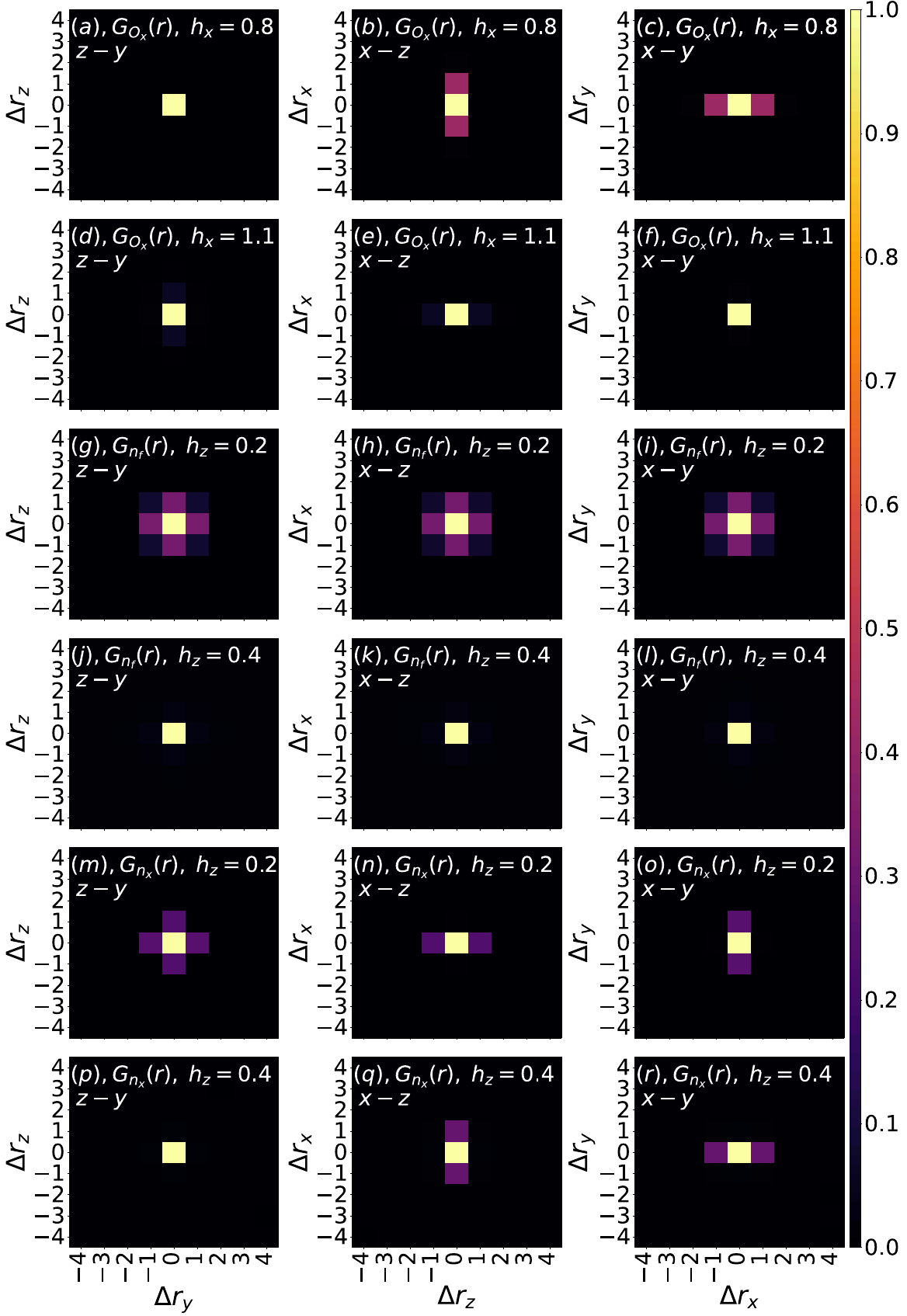}
	\caption{The real-space correlation function of lineons $C_{O_{x}}(r)$, fractons $C_{n_{f}}(r)$, and planons $C_{n_{x}}(r)$. Panels (a-c) [(g-i) and (m-o)] are in the fracton phase with $h_{x}=0.8$ [$h_{z}=0.2$], while the others are in the paramagnetic phase with $h_{x}=1.1$ (d-f) and $h_{z}=0.4$ [(j-l) and (p-r)]. The first, second, and third column describe the correlation function along the $y-z$, $x-z$ and $x-y$ planes, respectively.}
\label{fig:fig4}
\end{figure}

To show   mobility constraints on subdimensional excitations, we first measure their real-space normalized correlation functions defined as
\begin{equation}
	\begin{aligned}
		C_{O_{x}}(\textbf{r})&=\frac{\langle O_{x,i}O_{x,i+\textbf{r}} \rangle-\langle O_{x,i}\rangle^2}{\langle O_{x,i}^2 \rangle-\langle O_{x,i}\rangle^2},\\
		C_{n_{f}}(\textbf{r})&=\frac{\langle n_{f,i}n_{f,i+\textbf{r}} \rangle-\langle n_{f,i}\rangle^2}{\langle n_{f,i}^2 \rangle-\langle n_{f,i}\rangle^2},\\
		C_{n_{x}}(\textbf{r})&=\frac{\langle n_{x,i}n_{x,i+\textbf{r}} \rangle-\langle n_{x,i}\rangle^2}{\langle n_{x,i}^2 \rangle-\langle n_{x,i}\rangle^2}\,.
	\end{aligned}
\end{equation}
As discussed in Sec.~\ref{sec:theo_pre}, the anisotropic properties of $C_{O_{x}}$ and $C_{n_{x}}$ are tightly related to their own mobility in each direction. In Fig.~\ref{fig:fig4}, we present the numerical results of these correlation functions on a $L=10$ system.

The lineon-lineon correlation function, $C_{O_{x}}(\textbf{r})$, is anisotropic in the fracton phase. Taking $O_{x}$ with $h_{x}=0.8$ in the fracton phase as an example, $C_{O_{x}}(\textbf{r})$ decays more slowly along the $x$-direction with the neighboring correlation $C_{O_{x}}(\textbf{x})=0.421(3)$, but fast along the $y$- and $z$-direction with $C_{O_{x}}(\textbf{y})=0.0003(6)$ and $C_{O_{x}}(\textbf{z})=0.00002(6)$, which is shown in Fig.~\ref{fig:fig4}(a-c). Such a property comes from the mobility constraint that lineons can only move along $x$-direction. And in the paramagnetic phase, the fracton order is completely suppressed and it will be inappropriate to  interprete $O_x$ as lineon density. Therefore, the restriction of the lineon mobility also disappears, and so does the anisotropic property. As a result, $C_{O_{x}}(\textbf{r})$ rapidly decays to zero along all directions, which is given in Fig.~\ref{fig:fig4}(d-f).

 \begin{figure*}[htp!]
	\centering
	\includegraphics[width=\textwidth]{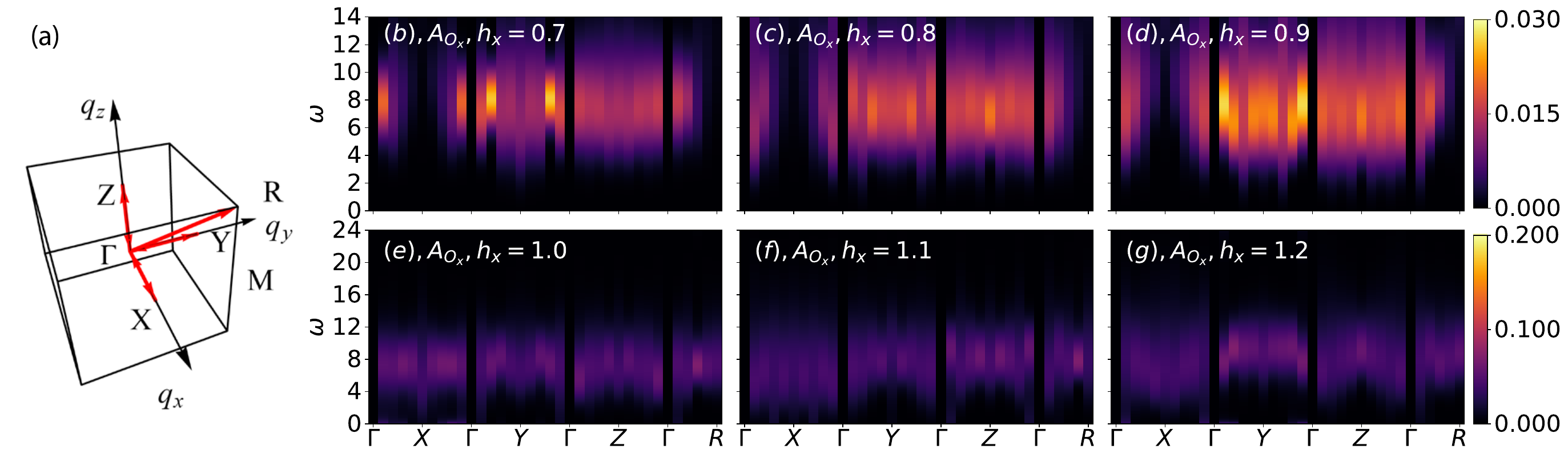}
	\caption{The lineon spectral function obtained from the QMC-SAC process and QA method. (a) the high-symmetry-path we plot in all three subdimensional excitations spectra. (b-d) are the spectrum $A_{O_x}$ in the fracton phase with $h_{x}$ ranging from $0.7$ to $0.9$, which are obtained from the QMC simulation by annealing from the exactly solvable point such that the configurations in each $h_x$ are fully thermalized. And (e-g), measured by scanning from the PLx point, tell $A_{O_x}$ in the trivial phase, in which $h_{x}$ comes from $1.0$ to $1.2$.}
	\label{fig:fig5}
\end{figure*}

\begin{figure*}[htp!]
	\centering
	\includegraphics[width=0.85\textwidth]{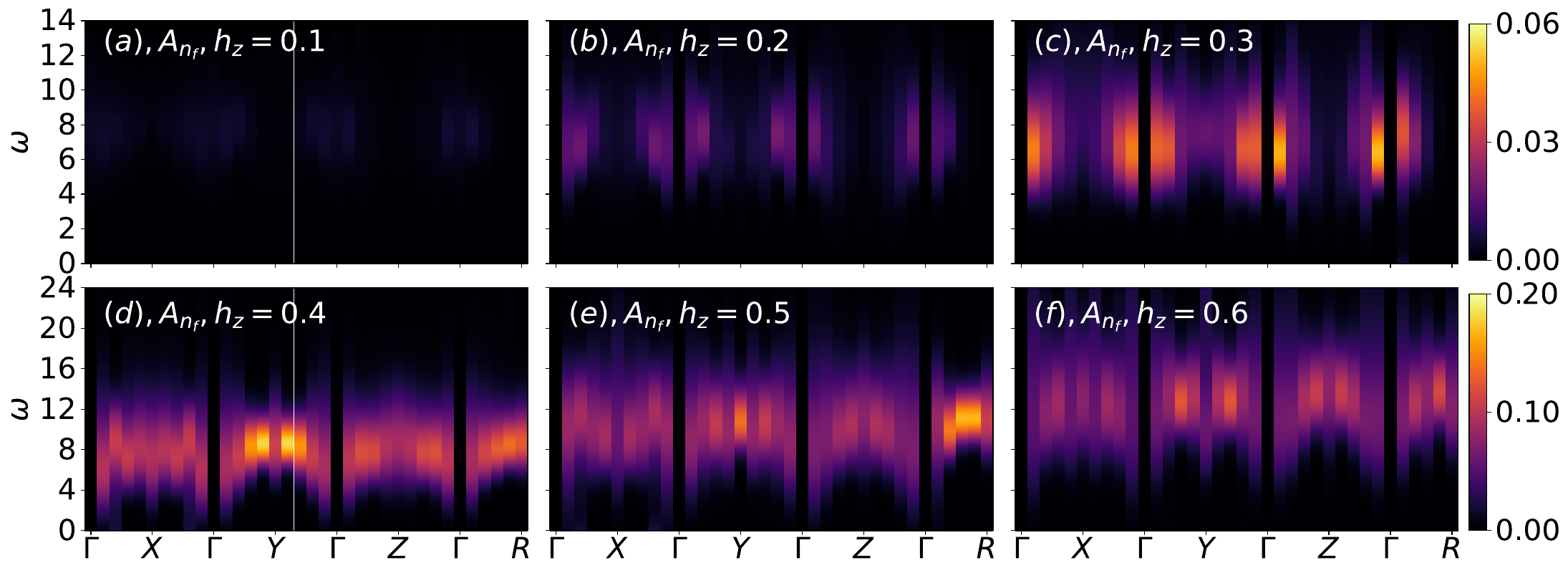}
	\caption{The fracton spectral function obtained from the QMC-SAC process and QA method. (a-c) are the spectrum $A_{n_f}$ in the fracton phase with $h_{z}$ ranging from $0.1$ to $0.3$, which is measured by using QMC simulation and quantum annealing from the exactly solvable point. And (d-f) tell $A_{n_f}$ in the trivial phase calculated by sweeping from the PLz point, in which $h_{z}$ changes from $0.4$ to $0.6$.}
	\label{fig:fig6}
\end{figure*}

\begin{figure*}[htp!]
	\centering
	\includegraphics[width=0.85\textwidth]{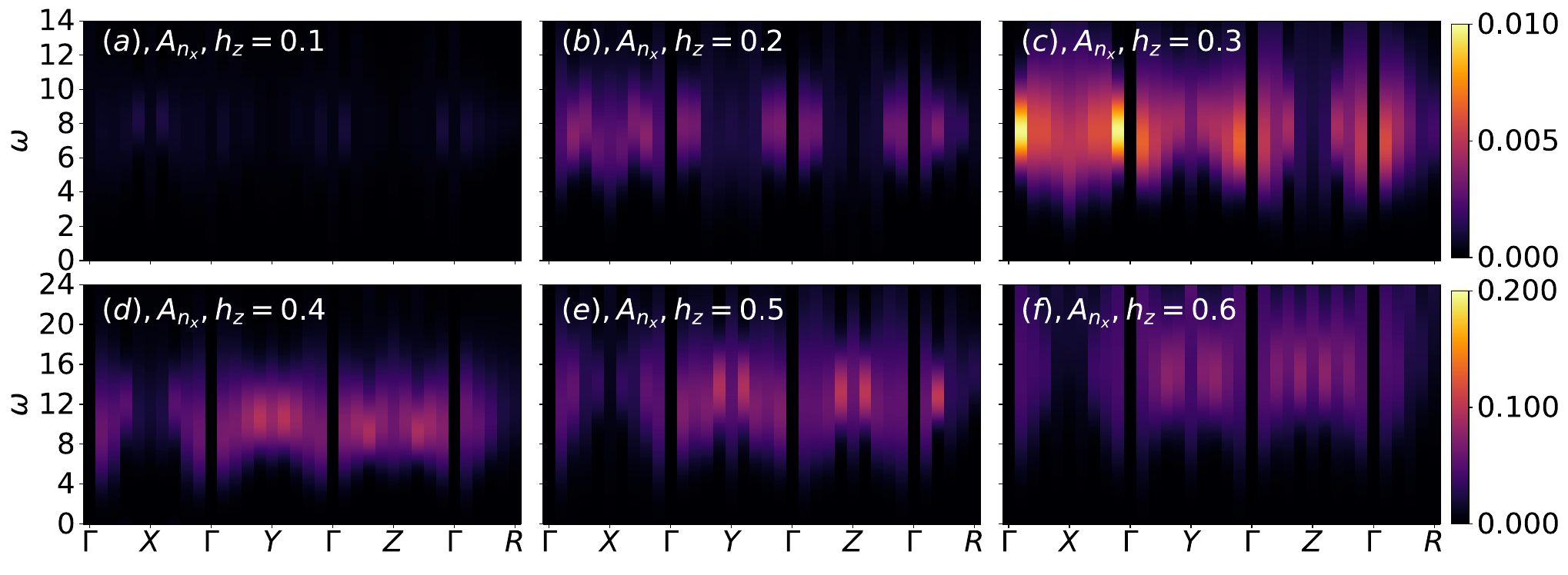}
	\caption{The planon spectral function obtained from the QMC-SAC process and QA method. (a-c) are the spectrum $A_{n_x}$ in the fracton phase with $h_{z}$ ranging from $0.1$ to $0.3$, which is measured by the QMC simulation with annealing from the exactly solvable point. And (d-f) tell $A_{n_x}$ in the trivial phase obtained by sweeping from the PLz point, in which $h_{z}$ ranges from $0.4$ to $0.6$.}
	\label{fig:fig7}
\end{figure*}
Although the fracton order can also be suppressed by the $h_{z}$ field, the real-space correlation function of planons, $C_{n_{x}}(\textbf{r})$, and fracton, $C_{n_{f}}(\textbf{r})$, behave differently from their 1-dimensional cousin due to their different mobility nature. For the planon, which is mobile in the $y-z$ plane, its correlation function $C_{n_{x}}$ decays slowly in both $y$- and $z$-direction with neighboring $C_{n_{x}}(\textbf{y})=0.251(6)$ and $C_{n_{x}}(\textbf{z})=0.238(8)$, but fast in the $x$-direction $C_{n_{x}}(\textbf{x})=0.007(2)$ [see Fig.~\ref{fig:fig4}(m-o)]. Such a restriction of mobility is broken when the model enters the paramagnetic phase. But, different from the lineon case, the planon operator $n_{x}$ can be viewed as a composite of two neighboring fracton operators $n_{f}$ in the $x$-direction. Consequently, $C_{n_{x}}(\textbf{r})$ remains anisotropic even in the paramagnetic phase. However, it slowly decays along its intrinsic structure direction ($x$-direction for $n_{x}$), which is shown in Fig.~\ref{fig:fig4}(p-r) with neighboring correlation $C_{n_{x}}(\textbf{x})=0.296(8)$. When it comes to fracton, it is immobile in all directions. This restriction leads to the isotropic correlation function in both fracton and paramagnetic phases, which looks featureless compared to the correlation data in the lineon channel. However, since the fractons are created at four corners by a membrane operator, $C_{n_{f}}(\textbf{x})=0.324(6)$ at the neighboring point in Fig.~\ref{fig:fig4}(g-i), which are $C_{n_{f}}(\textbf{x})=0.020(1)$ in the paramagnetic phase [Fig.~\ref{fig:fig4}(j-l)].

\subsection{Dynamic structure factors}
Beside the real-space correlation function, we also investigate their corresponding dynamic structure factors, i.e., spectral functions, $A_{O_x}(\mathbf{q},\omega)$, $A_{n_f}(\mathbf{q},\omega)$ and $A_{n_x}(\mathbf{q},\omega)$ obtained from the SAC upon the $G_{O_x}(\mathbf{q},\tau)$, $G_{n_f}(\mathbf{q},\tau)$ and $G_{n_x}(\mathbf{q},\tau)$ in Eqs.~\eqref{eq:eqOx}, \eqref{eq:eqnf} and \eqref{eq:eqnx}, which respectively correspond to the lineon, fracton, and planon channels. These results are presented in Figs.~\ref{fig:fig5}, \ref{fig:fig6} and \ref{fig:fig7}, where $L=10$. These spectral functions are plotted along the momentum path $\Gamma(0,0,0)\rightarrow X(\pi,0,0)\rightarrow\Gamma(0,0,0)\rightarrow Y(0,\pi,0)\rightarrow\Gamma(0,0,0)\rightarrow Z(0,0,\pi)\rightarrow\Gamma(0,0,0)\rightarrow R(\pi,\pi,\pi)$, as illustrated  in Fig.~\ref{fig:fig5}(a).

In Figs.~\ref{fig:fig5}, \ref{fig:fig6} and \ref{fig:fig7}, the first rows are in the fracton phase, with external fields $h_{x}=0.7-0.9$ in Fig.~\ref{fig:fig5}(b-d) and $h_{z}=0.1-0.3$ in both Fig.~\ref{fig:fig6}(a-c) and \ref{fig:fig7}(a-c). Meanwhile the second rows show the paramagnetic phase with $h_{x}=1.0-1.2$ [Fig.~\ref{fig:fig5}(e-g)] and $h_{z}=0.4-0.6$ [Fig.~\ref{fig:fig6}(d-f) and \ref{fig:fig7}(d-f)]. From the results, we see that, for any given momentum point, the spectral functions in all three channels do not have signals when frequency is below   a finite value. Therefore, the spectral functions have finite gaps in the fracton phase, meaning that the density fluctuations of all three kinds of subdimensional excitations must cost finite energy. This   is expected, since all subdimensional excitations have finite single-particle gaps. Furthermore, in Figs.~\ref{fig:fig5}(b-d), \ref{fig:fig6}(a-c), and \ref{fig:fig7}(a-c), we find that, upon increasing external fields, quantum fluctuations get stronger and stronger, which causes more and more pronounced spectral signals (quantified by brightness in the $\omega-k$ plane) in the fracton phase. And in the paramagnetic phase shown in the second rows of Figs.~\ref{fig:fig5}, \ref{fig:fig6}, and \ref{fig:fig7},  the gaps for density fluctuations become larger and larger upon increasing external fields.  We must stress that, since subdimensional particles are no longer well-defined   excitations in the trivial  paramagnetic phase, rigorously speaking, the spectral functions should no longer be interpreted as measurement of density fluctuations of subdimensional particles.

Let us first focus on the spectral function of lineons. The peak energy (i.e., the frequency of the brightest signals) of the spectral function of lineons  is approximately given by $\Delta\approx8$ in both the fracton phase [Fig.~\ref{fig:fig5}(b-d)] and the paramagnetic phase [Fig.~\ref{fig:fig5}(e-g)]. However, the profile of the spectral function in the fracton phase clearly shows a dispersion behavior along the $\Gamma\rightarrow X$, while it is flat in the paramagnetic phase. Theoretically, this dispersive behavior of density fluctuations can be traced back to the free mobility of lineons along $x$-direction.
Moreover, in the paramagnetic phase, the profile of the spectral function is flat over the whole momentum space, but  its peak energy   becomes larger and larger upon  increasing $h_{x}$ from Fig.~\ref{fig:fig5}(e) to (g).

\begin{figure}[htp!]
	\centering
	\includegraphics[width=\columnwidth]{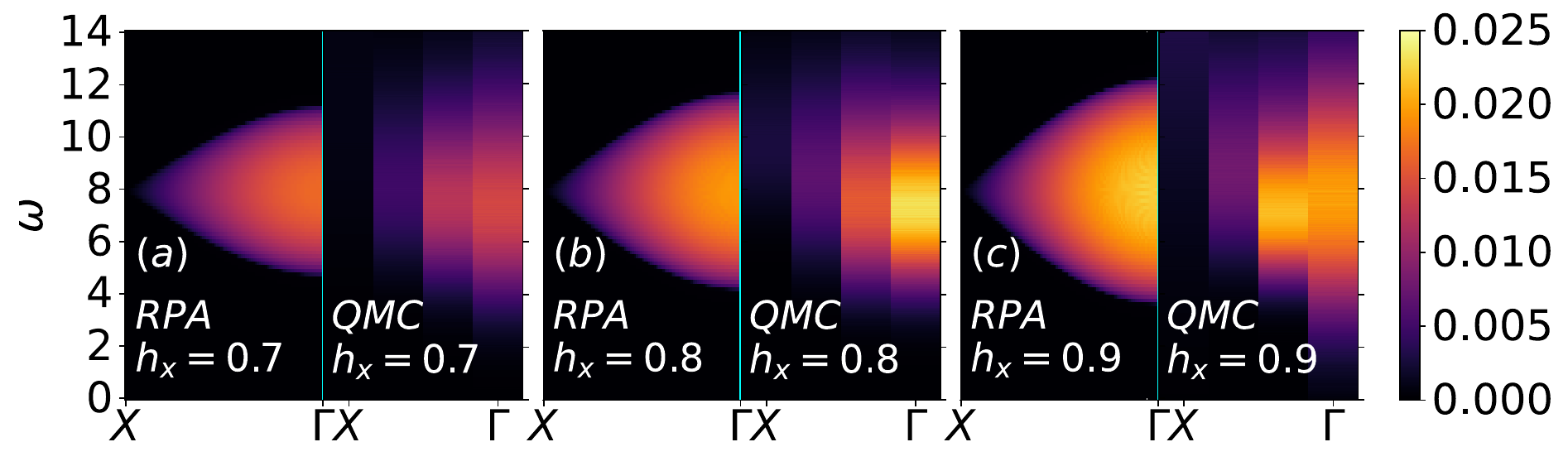}
	\caption{Comparison between the dynamic structure factors in the lineon channel obtained by mean-field+RPA method   and  the QMC process. The momentum range is ($X \rightarrow \Gamma$) along which lineons are dispersive. The RPA results are carried out in with coupling constant $J=14$. Fig.~\ref{fig:DSF}(a-c) are in the fracton phase with the external field $h_{x}$ raising from $0.7$ to $0.9$. Here, the QMC results are simulated at system size $L=6$.
	%The dashed cyan lines on the right hand side in (a), (b) and (c) are the peak location by fitting the imaginary correlation function $G(\textbf{q},\tau)$ with $exp\left(-\Delta\tau\right)$. Here, $\Delta$ refers to the energy gap.
	}
	\label{fig:DSF}
\end{figure}

Next, we perform a mean-field analysis on the spectral functions of lineons. Since the kinetic energy of lineons along dispersionless directions is strictly suppressed such that interaction effect is no longer negligible, which motivates us to  focus on dispersive directions. Without loss of generality, we consider lineons mobile along $x$ direction. At first, we regard $O_{x,i}$ as a density operator of lineons on site $i$. Then, we need to notice that the ground state of X-cube model perturbed by Zeeman fields is no longer a ``vacuum'' for lineons but contains finite density of lineons.  Furthermore, by noticing that lineons are hard-core and mobile within one-dimensional subspace, and we expect the difference between the exchange phases of hardcore bosons and fermions to have little influence when there is few particles in 1D (thus particles can hardly exchange positions), it is phenomenologically reasonable to model lineons as 1D fermions with both positive and negative energy spectra where all states with negative energy are filled. Therefore,   we propose a density-density correlation function \cite{Ehrenreich_Cohen1959,adler1962} of lineons of the following form to fit the QMC results of spectral functions:
\begin{align}\chi_0(q,\omega) = \sum_{\nu, \nu^\prime} \sum_k \frac{1}{V} \frac{n_F(\xi^{\nu^\prime}_{k+q}) - n_F(\xi^{\nu}_k)}{\omega + i\eta - (\xi^{\nu}_k - \xi^{\nu^\prime}_{k+q})}\,.
\end{align}
Here, $\xi_{k}^\nu$ is a gapped dispersion relation. The indices $\nu,\nu^\prime=\pm$ and we require that $\xi_{k}^+ = -\xi_{k}^-$ and $\xi_{k}^+ > 0$. $n_F(\xi) =   ({e^{\beta \xi} + 1})^{-1}$ is the Fermi-Dirac distribution, and $\beta$ is the inverse temperature. $V$ is the volume of the system and $\eta$ is an infinitesimal real number. Besides, compared with the result in Ref.~\cite{Ehrenreich_Cohen1959,adler1962}, a $k$-dependent matrix element has been omitted, since here only the qualitative tendency is of concern. Then, at zero temperature,   $\chi_0(q,\omega)$   can be reduced to
 $\chi_0(q,\omega) \sim {L}^{-1}\sum_{k}  [ ({\omega + i\eta - \xi_{k}^+ - \xi_{k+q}^+})^{-1} - ({\omega + i\eta + \xi_{k}^+ + \xi_{k+q}^+})^{-1}]$, where the second term can be safely neglected when $\omega > 0$. Since the lineons we consider here are only mobile along $x$ direction, the volume $V$ reduces to the linear size $L$  upon summing over $k_y$ and $k_z$.    To proceed further,  we   assume the lineon dispersion to be:
  $  \xi^+_{k} = 4 - (4-\delta) \cos(k_{x})\,,
  $  where the energy gap $\delta = 4 - 2 h_x - \frac{1}{2} h_x^2 + \frac{1}{8} h_x^3 - \frac{53}{192} h_x^4 + \frac{973}{4608} h_x^5$ is taken from the perturbative result in  Ref.~\cite{Quantum2020Muhlhauser}. By further phenomenologically considering a RPA approximation, i.e.,
  \begin{align}
   \chi_{RPA}(q,\omega) = \frac{\chi_0(q,\omega)}{1 + J\chi_0(q,\omega)}
   \end{align}
    to effectively include the interaction effect ($J$ is a coupling constant), the dynamic structure factor (spectral function) can be obtained as $A(q,\omega) = -\text{Im} \chi_{RPA}(q,\omega)$. Here, we set parameters $\eta = 0.01, L=2000, J = 14$. We compare the RPA results and QMC results in Fig.~\ref{fig:DSF}, which shows a qualitative consistency between the two approaches. The QMC results in Fig.~\ref{fig:DSF} are carried out on a $L=6$ system also annealing from the exactly solvable point but with more bins in the QMC measurement, which help us to reduce the error in $G_{O_{x}}$ and solve out a high-quality spectral function $A_{O_{x}}$.

While a more standard slave-particle mean-field treatment \cite{WenRMP,auerbach2012interacting} is desirable, we leave it to future investigation.  As lineons, fractons, and planons are localized along some certain directions, it will be also interesting to probe such localizations. One of possible ways is to compute complementary entanglement properties from both real space cut and momentum space cut via techniques and observables in e.g., Refs.~\cite{mondragonshemCharacterizing2013,Chen:2021akh,leePositionmomentum2014,chenEntanglement2020}, which we also leave to future investigations.

After analyzing the spectral function for lineons, let us move to the spectral functions of fractons and planons. Figs.~\ref{fig:fig6} and \ref{fig:fig7} respectively show the spectral functions of fracton and planon excitations with $h_{z}$ changing from $0.1$ to $0.6$ on a $L=10$ system, respectively. Since $h_{z}$ term does not commute with the $A_{c,i}$, it leads to a fluctuation of $A_{c,i}$ and causes the appearances of fractons and planons.
For small perturbation ($h_{z}\leq0.3$), the X-cube model stays in the fracton phase. Figs.~\ref{fig:fig6}(a-c) shows the fracton spectral function in the fracton phase, and Figs.~\ref{fig:fig7}(a-c) are that of planon. In panel (a) of Figs.~\ref{fig:fig6}-\ref{fig:fig7}, the perturbation from $h_{z}$ is too weak and the dynamic signal of the fracton and planon is hard to be observed. By increasing $h_{z}$, the stronger fluctuations in $n_{f}$ and $n_{x}$ make their corresponding spectral signal more significant from panel (a) to (c). The profiles of spectral functions of both fractons and planons are flat with peak energy $\Delta\approx8$, which is approximately four times of single fracton energy~\cite{Quantum2020Muhlhauser}. When $h_{z}\geq0.3$, the X-cube model enters the trivial paramagnetic phase. In this case, the profiles of spectral functions of both fracton and planon channels remain flat but the peak energy is becoming larger with stronger perturbation, which are presented in Figs.~\ref{fig:fig6}(d-f) and \ref{fig:fig7}(d-f).

In summary, the most significant feature in the fracton phases in Fig.~\ref{fig:fig5}-\ref{fig:fig7} is that the dispersion behaviors of these spectral functions show strong anisotropy corresponding to their mobility restriction. For example, in the Fig.~\ref{fig:fig5}(b-d), the profile of the spectra of lineon in the fracton phase clearly shows a dispersive behavior along the $\Gamma\rightarrow X$, while it is flat in the paramagnetic phase (Fig.~\ref{fig:fig5}(e-g)). Meanwhile, with the help of the mean-field+RPA method, we can reproduce the profile of the lineon spectra along the dispersive direction as well. Besides, for fracton spectra as shwon in Fig.~\ref{fig:fig6}, we see no obvious anisotropy, that is also consistent with the totally immobile feature of fractons. As a result, we believe the anisotropic behavior is a characteristic feature of Type-I fracton orders, where partially mobile excitations exist.

\section{Summary and outlook}
\label{sec:discussion}

Before closing this paper, let us briefly summarize  our  numerical findings.
In the lineon channels, by means of the quantum fluctuations from the $h_{x}$ field,   lineons are excited, whose particle number becomes finite and dense in the ground state. In the fracton phase, the real-space lineon-lineon correlation shows a significant value on the neighboring points along its one-dimensional mobile direction, e.g., the $x$-direction for $O_{x}$. Meanwhile, the peak profile of the spectral function of $O_{x}$ also presents a dispersive behavior along the $\Gamma\rightarrow X$ direction. With the help of the mean-field theory and RPA method by assuming an anisotropic dispersion relation of lineons, we qualitatively recover the QMC results of the spectral function of lineons, which  demonstrates  an exotic connection between  spectral functions and   mobility constraints on    dynamics of lineons.
By comparing with the results in the trivial paramagnetic phase, these anisotropic properties can serve as an experimental signature of the fracton order.

In the fracton and planon channels, under the perturbation of $h_{z}$ field, the fluctuation of $A_{c,i}$ terms leads to the appearance of both fracton and planon excitations. The fracton-fracton correlation function is isotropic in both fracton and paramagnetic phase, which is due to its completely immobile nature. However, in the plannon case, its correlation function behaves anisotropically in both phases but for different reasons. It is due to its mobility constraint in the fracton phase but the intrinsic dipole-like structure in the paramagnetic case. Such a difference leads to a significant geometrical rotation of these anisotropic features, which could be a very experimental signal for the purpose of identifying the type-I, i.e., X-cube fracton order. And, for spectral functions, the spectral peak energies in both these channels in the fracton phase are non-sensitive against the increasing $h_{z}$ but become  sensitive in the paramagnetic case.

 In conclusion, with the external Zeeman-like magnetic fields pushing the X-cube model away from its exactly solvable point, we have improved QMC methods to overcome the glassy Hilbert space, and computed both real-space correlation functions and dynamic structure factors in three subdimensional excitation channels in both the fracton and the paramagnetic phases through the method of QMC+SAC in a wide parameter space. We have found how   mobility constraints on local dynamical properties of subdimensional excitations is correlated to the existence of fracton topological order. This fact is nothing but  the  exotic interplay of topology (global topological order) and geometry (local dynamical properties) in fracton orders. As our numerical results show novel dynamical features that are yet to be fully understood, we expect our numerical results will stimulate more efforts in both theoretical and numerical studies in fracton physics. While the X-cube Hamiltonian with magnetic fields looks intricate, it will be interesting and challenging to analytically study correlation   and spectral functions in the theoretical framework of slave-particle (projective construction) techniques. Our  numerical results also set a further interesting question about the connection between the single-particle dispersions of the lineon, fracton and planon channels and their density spectral functions~\cite{Quantum2020Muhlhauser}.
 Finally, as spectroscopy measurements have been widely performed in strongly-correlated materials, our numerical results on dynamical signature will  be helpful in experimental identifications of fracton orders in spectroscopy measurements such as neutron scattering and nuclear magnetic resonance.

%\section*{Acknowledgement}
\acknowledgements
C.K.Z., Z.Y., and Z.Y.M. acknowledge support from the Research Grants Council of Hong Kong SAR of China (Grant Nos. 17303019, 17301420, 17301721 and AoE/P-701/20), the GD-NSF (no. 2022A1515011007), the K. C. Wong Education Foundation (Grant No. GJTD-2020-01) and the Seed Funding “Quantum-Inspired explainable-AI” at the HKU-TCL Joint Research Centre for Artificial Intelligence. M.Y.L. and P.Y. thank Xiao-Xiao Zhang, Yi-Zhuang You, Yao Zhou, Chao Gao,  Chao-Ming Jian, and Gang Chen for helpful discussions. M.Y.L. and P.Y. were supported by Guangdong Basic and Applied Basic Research Foundation under Grant No.~2020B1515120100, NSFC Grant (No.~11847608 \& No.~12074438). We thank the Computational Initiative at the Faculty of Science and HPC2021 system under the Information Technology Services at the University of Hong Kong, and the Tianhe-II platform at the National Supercomputer Center in Guangzhou and the Beijing PARATERA Tech Co., Ltd. for providing HPC resources that have contributed to the research results reported within this paper. The work was performed in part on resources provided by the Guangdong Provincial Key Laboratory of Magnetoelectric Physics and Devices (LaMPad). The work was performed in part on resources provided by the Guangdong Provincial Key Laboratory of Magnetoelectric Physics and Devices (LaMPad).

\bibliography{ref}

\newpage
\appendix
\section{Numerical method}
\label{sec:numerical method}
Here, we perform our simulation for the X-cube model using QMC method. Taking the $\sigma_{x}$ basis as an example, we first rewrite the Hamiltonian as

\begin{equation}
	\begin{aligned}
		H&=-\sum_{i} H_{A{c,i}}-\sum_{v,i} H_{B_{v,i}}-\sum_{i} H_{h_{x,i}}-\sum_{i} H_{h_{z,i}}+\sum H_{C},
	\end{aligned}
	\label{eq:eq2}
\end{equation}
with $H_{A_{c,i}}=K(A_{c,i}+3I_{A})$, $H_{B_{v,i}}=\Gamma(B_{v,i}+I_{B})$, $H_{h_{x,i}}=h_{x}(\sigma^{x}_{i}+3I_{2})$, and $H_{h_{z,i}}=h_{z}(\sigma^{z}_{i}+I_{2})$. And, $H_{C}=3KI_{A}+\Gamma I_{B}+3h_{x}I_{2}+h_{z}I_{2}$, where $I_{A}$, $I_{B}$ and $I_{2}$ are identity matrix of order $2^{12}$, $2^4$ and $2$. Following the framework of QMC method, all of the none-zero elements in the Hamiltonian can be read as $\langle H_{A{c,i}} \rangle=2K$ or $4K$, $\langle H_{B_{v,i}} \rangle=\Gamma$, $\langle H_{h_{x,i}} \rangle=2h_{x}$ or $4h_{x}$, and $\langle H_{h_{z,i}} \rangle=h_{z}$.

As mentioned in Sec.~\ref{sec:numerical calculation}, caused by the four ($B_{v,i}$) and twelve-spin ($A_{c,i}$) interactions, the efficiency of the cluster algorithms may be slowed down by the over-raptly extending cluster\cite{Stochastic2003Sandvik}.
 To make our simulation more effective, we first modify the typical cluster update algorithm to restrict the cluster extending. Then, we apply both the modified cluster update and the local update in our simulations. Moreover, at the exactly solvable point, we also deduce a helpful initial configuration according to an equilibrium configuration of a smaller system size simulation. Finally, due to the first-order phase transition, we also need to utilize the QA algorithm. Our simulation scans from the exactly solvable point in the fracton phase measurement, and from the PLz(x) point in the paramagnetic phase measurement (see Fig.~\ref{fig:fig3}).

\subsection{Diagonal update}
\label{sec:appA1}
The diagonal update is about inserting and removing the operators. In the inserting process, the whole number of possible insert positions are
\begin{equation}
	N=
	\begin{cases}
		&N_{c}+N_{v}+2N_{l} \quad h_{z}\neq0 \quad h_{x}\neq0\\
		&N_{c}+N_{v}+N_{l} \quad h_{z}\neq0 \quad or \quad h_{x}=0 \\
		&N_{c}+N_{v}+N_{l} \quad h_{z}=0 \quad or \quad h_{x}\neq0 \\
		&N_{c}+N_{v} \quad h_{z}=0 \quad h_{x}=0\\
	\end{cases}
	\label{eq:eq5}
\end{equation}
Here $N_{c}$ is the total number of the cubes and $N_{v}$ for the crosses. And $N_{l}$ is the number of the spins. We prefer a position by randomly picking a number $N_{x}$ in $N$, which determine the operator type would be inserted. If $1\leq N_{x}\leq N_{c}$, we propose to insert a $H_{A{c,i}}$ operator at the $N_{x}$ cube. When $N_{c}+1\leq N_{x}\leq N_{c}+N_{v}$, a $H_{B_{v,i}}$ operator is suggested to inserted at the $N_{x}$ crosses. And, if $N_{c}+N_{v}+1\leq N_{x}\leq N_{c}+N_{v}+N_{l}$ and $h_{x}\neq0$, a $H_{h_{x,i}}$ operator is applied to corresponding spin. Moreover, when $N_{c}+N_{v}+1\leq N_{x}\leq N_{c}+N_{v}+N_{l}$ and $h_{x}=0$, or $N_{c}+N_{v}+N_{l}+1\leq N_{x}\leq N_{c}+N_{v}+2N_{l}$, we propose to add a $H_{h_{z,i}}$ operator on the $N_{x}$ spin.

After an identity operator is encountered, such insertion of the operator $H_{D}$ would be accepted with probability
\begin{equation}
	\begin{aligned}
		P_{add}{(H_{D})}&=\frac{\beta N \langle H_{D}\rangle}{m-n},
	\end{aligned}
	\label{eq:eq6}
\end{equation}
with $m$ for the length of the configuration and $n$ for the operator number. Noted that the value of $\langle H_{D}\rangle$ not only depends on the type of operator, but also the spin states at the corresponding position.

In the removing case, we sweep the configuration and remove the diagonal operator $H_{D}$ with probability
\begin{equation}
	\begin{aligned}
		P_{rem}{(H_{D})}&=\frac{m-n+1}{\beta N \langle H_{D}\rangle}.
	\end{aligned}
	\label{eq:eq10}
\end{equation}

\subsection{Off-diagonal update}
\label{sec:appA2}
When it comes to the off-diagonal update process, both the local update and the modified cluster update are applied in our simulation. Firstly, there are two kinds of operators in the configuration space, which are the pure diagonal operator ($H_{A{c,i}}$ and $H_{h_{x,i}}$) and the quantum operator ($H_{B_{v,i}}$ and $H_{h_{z,i}}$). Due to constant term in Eq.\ref{eq:eq2}, the diagonal element in $H_{B_{v,i}}$ and $H_{h_{z,i}}$ are no-zero. Thus, in the configuration, the quantum operator can be both diagonal and off-diagonal.

\begin{figure}[htp!]
	\centering
	\includegraphics[width=\columnwidth]{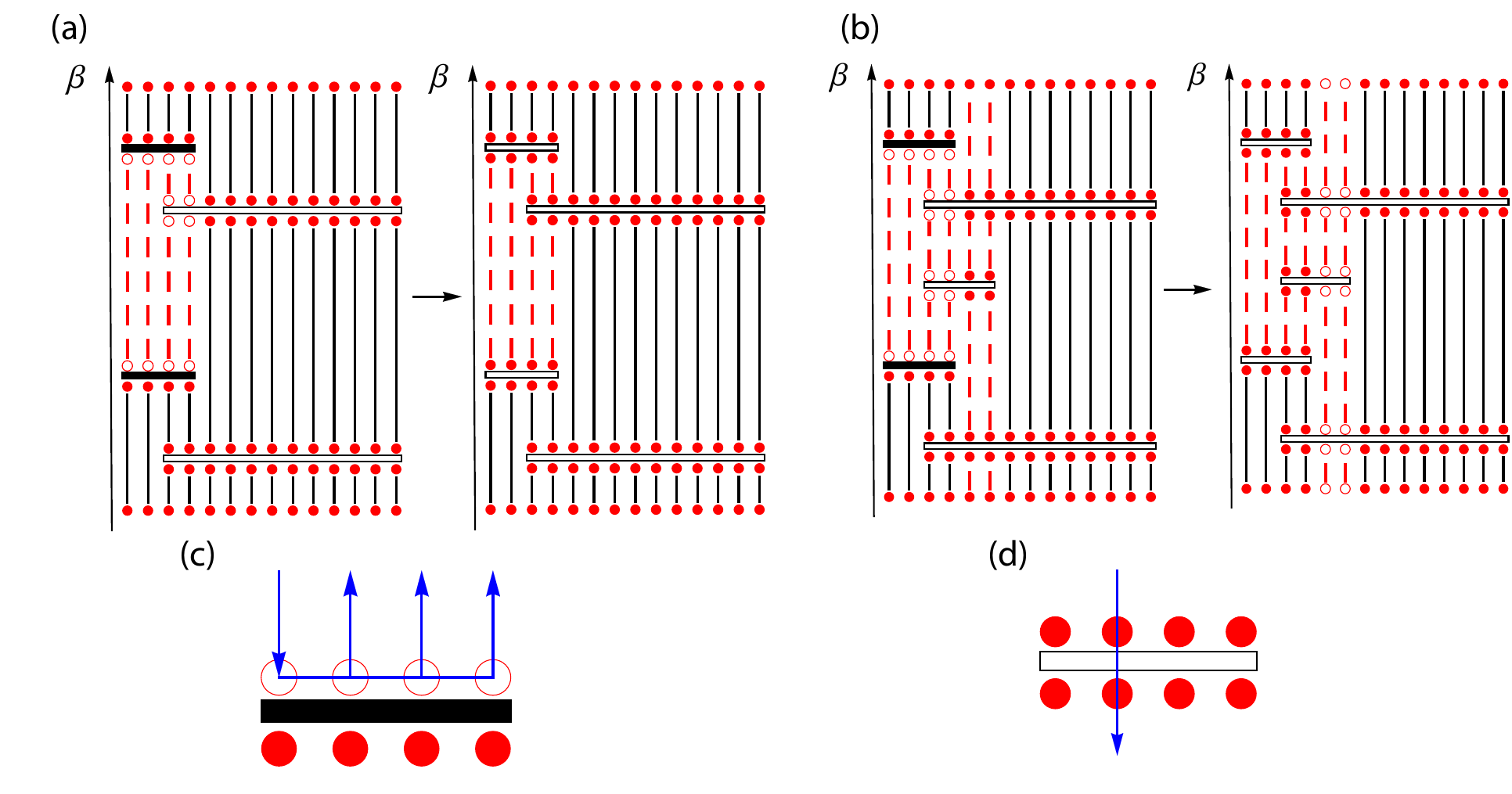}
	\caption{Schematic diagrams of the off-diagonal update. The red dashed line is the region suggested to flip the spin. Under the Metropolis process, its acceptance probability is $P_{fl}=min\left(1,\frac{W_{new}}{W_{old}}\right)$. (a) refers to the local update and (b) is the cluster update. And (c) and (d) describe two different way when a cluster line meets an operator.}
	\label{fig:axfig1}
\end{figure}

In the local update process, a leg of a quantum operator ($H_{B_{v,i}}$ and $H_{h_{z,i}}$) is selected randomly in the configuration. Then, starting from this leg, we create all the update-lines of this vertex and go along the imaginary time until finding another operator acting on the same position (see Fig.~\ref{fig:axfig1}(a)). The spins between these two operators are proposed to be flipped (the red region in the Fig.~\ref{fig:axfig1}(a)).

For the cluster update, also starting from a randomly picked-up vertex leg of a quantum operator, the cluster is built following the rules as below. (1) When the cluster building line meets a pure diagonal operator ($H_{A{c,i}}$ and $H_{h_{x,i}}$), it would go through the operator directly as Fig.~\ref{fig:axfig1}(d) shows. (2) When the cluster building line meets a quantum operator ($H_{B_{v,i}}$ and $H_{h_{z,i}}$), it would evolve in two different ways. This line can go through the operator directly as Fig.~\ref{fig:axfig1}(d). Or it will be reflected from the other vertex leg as Fig.~\ref{fig:axfig1}(c). In each cluster update process, we pick $10\%$ of these quantum operators randomly and treat them in the Fig.~\ref{fig:axfig1}(c) way in the cluster constructing process while the others in the Fig.~\ref{fig:axfig1}(d) ways. Within this treatment, the cluster extending would be slowed down, which make the cluster update more effective. And it is worth to note that this cluster building process would turn back into the typical cluster update with treat each quantum operators in the Fig.~\ref{fig:axfig1}(c) way. Finally, the spins including in the cluster (the red region in the Fig.~\ref{fig:axfig1}(b)), are suggested to be flipped.

In both of these updates, the spins included in the red region would be flipped with the acceptances probability given by
\begin{equation}
	\begin{aligned}
		P_{fl}&=min\left(1,\frac{W_{new}}{W_{old}}\right),\\
		\frac{W_{new}}{W_{old}}&=2^{n^{+,new}-n^{+,old}}.
	\end{aligned}
	\label{eq:eq12}
\end{equation}

Here, $n^{+,new (old)}$ are the number of diagonal operators with $\langle H_{Ac(h_{x})}\rangle=4K(h_{x})$ in the new (old) configuration. It means that the weight ratio depends on the number of the overlap-value-changing diagonal operator.

\subsection{Deducing initial configuration}
\label{sec:appA3}
In the Monte Carlo simulation, the final results should be independent of the initial configuration. However, a better choice of the initial configuration is helpful to reduce the computational cost. To find such a better choice, with the concept of super-resolution, we deduce an initial configuration of $(fL)^3\beta$ system from an equilibrium $L^3\beta$ configuration. And here, in our case, the initial configuration of $L=10$ is deduced from the $L=2$ result with $f=5$.

Under the framework of the SSE method, the partition function is expanded as
\begin{equation}
	\begin{aligned}
		Z&&=\sum_{\{\alpha\}_n}\sum_{S_{m}}\frac{\beta^n(m-n)!}{m!}\langle\alpha_0|\prod^{n}_{i=1} H_i|\alpha_0\rangle.
	\end{aligned}
	\label{eq:eq4}
\end{equation}
This expansion constructs a configuration space for the Monte Carlo sampling. Such a configuration keeps two important features, the operator string $\prod^{n}_{i=1} H_i$ and the initial state $|\alpha_0\rangle$. Finding a better initial configuration by super-resolving is to find a mapping from the configuration of $L^3\beta$ to that of $(fL)^3\beta$, which would make fewer mutations in physic values observing from the QMC process, such as energy density and magnetization. To do so, we discuss the operator string and the initial state respectively.

\begin{figure}[htp!]
	\centering
	\includegraphics[width=\columnwidth]{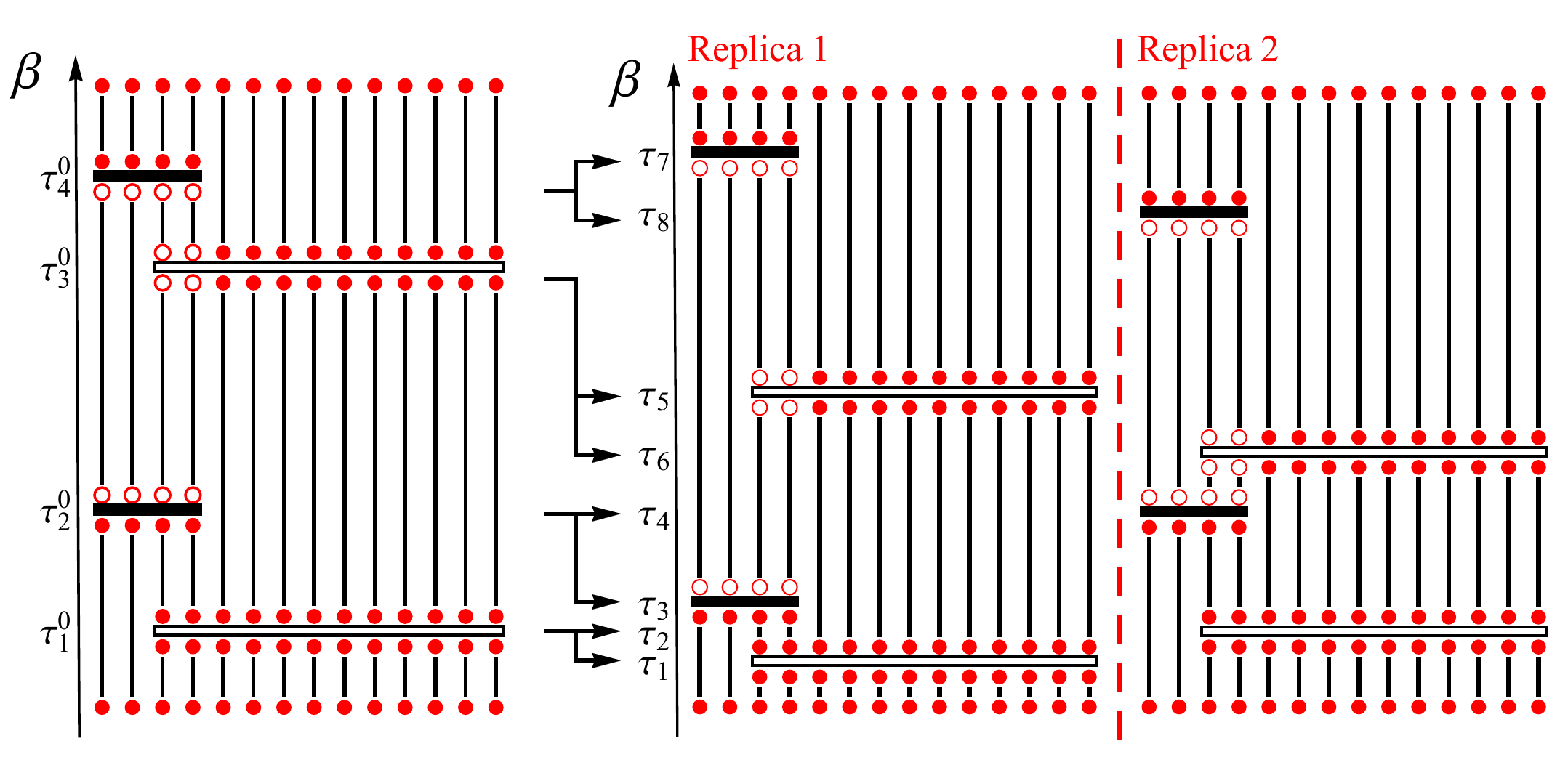}
	\caption{Schematic diagrams of the super-resolution process. The right figure denotes the original configuration of the small system, while the left figure is the extended configuration for a larger system. Operator $\tau^{0}_{i}$ is mapped to new operators $\tau_{f^3(i-1)+1}$ to $\tau_{f^3i}$ for each replica.}
	\label{fig:axfig2}
\end{figure}

For the operator string, with the periodic boundary condition, the super-resolution process can be viewed as making $f^3$ replicas but shifting the location of the operators. Firstly, the operators in the original configuration are marked as $\{\tau^{0}_{i}\}$, and the length of the configuration is extended $m\rightarrow f^3m$. Then, we pick up $f^3n$ positions in $f^3m$ randomly, which are marked as $\{\tau_{j}\}$. And we place the $\tau^{0}_{i}$ operator on the positions $\tau_{f^3(i-1)+1}$ to $\tau_{f^3i}$ for each replicas. For instance, in Fig.~\ref{fig:axfig2}, the $\tau^{0}_{1}$ operator is replaced on the positions $\tau_{1}$ and $\tau_{2}$. In this way, we keep the fact that the operators should appear uniformly in the configuration spaces. When it comes to the initial state, noted that the ground state $|GS\rangle$ of the X-cube model is
\begin{equation}
	|GS\rangle=\prod\frac{1+B_{v,i}}{2}\prod\frac{1+A_{c,i}}{2}|\uparrow\uparrow...\uparrow\rangle,
\end{equation}
where $|\uparrow\uparrow...\uparrow\rangle$ denote the fully polarized state\cite{Cage-Net2019Prem,isotropic2017vijay,Quantum2020Muhlhauser}. Therefore, we prefer the fully polarized state in the chosen basis as the new initial state of the larger system size configuration. Starting from this new configuration, with further Monte Carlo steps ($\geq10^4$), a new equilibrium configuration will be reached. In our work, we applied such a process in the simulation at the exactly solvable point.

Then, due to the first-order phase transition caused by $h_{x}$ and $h_{z}$, our simulation is also required the QA process to achieve a faster convergence to the ground state under the external Zeeman-like magnetic fields, in which the quantum parameter, $h_{x(z)}$, would be exactly slowly changed and the configuration from the last parameter result would be set as an initial configuration for QMC simulation at the next parameter. Our simulation in the fracton phase of the X-cube model scans from the exactly solvable point with an annealing step $\Delta h_{x(z)}=0.01$ and over $2\times10^5$ Monte Carlo steps at each annealing step\cite{kadowaki1998quantum,Car2002,yan2022targeting}. And the measurements in the paramagnetic phase are from PLz(x) points in Fig.~\ref{fig:fig3} with the same annealing step.

\begin{figure}[htp!]
	\centering
	\includegraphics[width=\columnwidth]{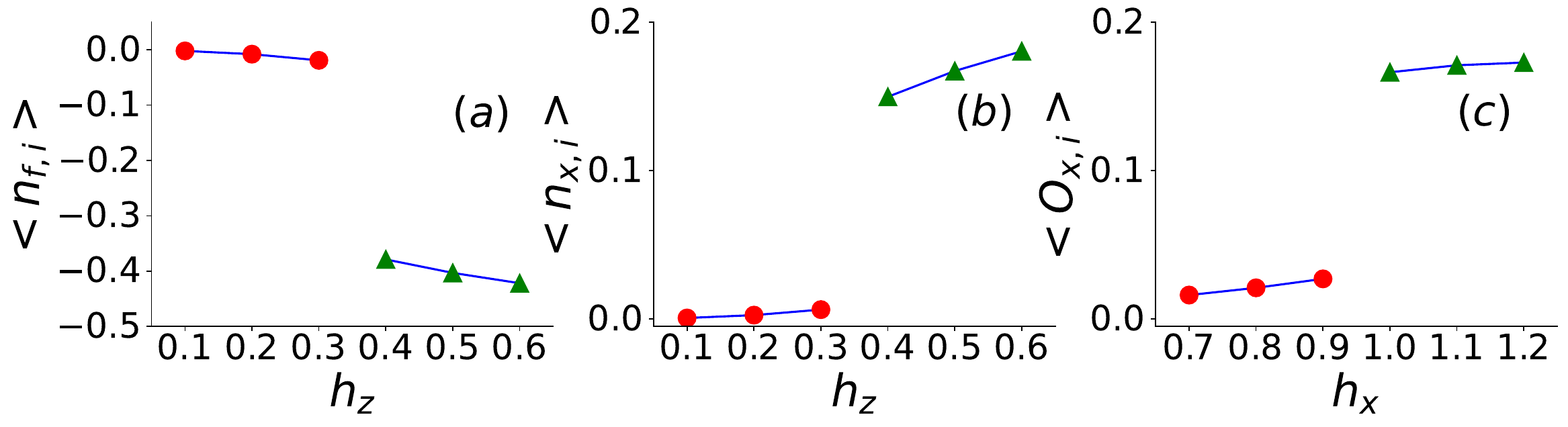}
	\caption{The values of the density operators change with various field. The red circles refer to the value in the fracton phase while the green triangle is in the trivial phase. (a) tells the density operator of fractons changing as the external Zeeman field increases. (b) is for the density operator of planons and (c) for lineons case.}
	\label{fig:axfig3}
\end{figure}

Moreover, we measure the density operators changing with the external Zeeman field, which are plotted in Fig.\ref{fig:axfig3}. In Fig.\ref{fig:axfig3}, the red circles are in the fracton phase while the green triangle are in the trivial phase. As it is shown, in the fracton phase, the density operators of all three kinds of subdimesional excitations are weak meaning that these subdimensional exicitations are hard to be created in the fracton phase. In contrast, these density operators have significant values in the trivial phase, which caused by the strong quantum fluctuation due to the external Zeeman fields. Meanwhile, the value of these density operators change discontinuously near the quantum critical point, which is due to the first order phase transition nature between the fracton and the trivial phase. 	

\setcounter{page}{1}
\setcounter{equation}{0}
\setcounter{figure}{0}
\renewcommand{\theequation}{S\arabic{equation}}
\renewcommand{\thefigure}{S\arabic{figure}}

\newpage

\end{document}